\documentclass[12pt]{article}
\newcommand{\be}{\begin{eqnarray}}
\newcommand{\ee}{\end{eqnarray}}
\newcommand{\ba}{\begin{eqnarray*}}
\newcommand{\ea}{\end{eqnarray*}}

\tiny\normalsize

\usepackage{graphicx} 
\usepackage{booktabs} 
\usepackage{amsmath}
\usepackage{mathtools}
\usepackage{color, colortbl}
\usepackage{tikz}
\usepackage{multirow}
\usetikzlibrary{fit,positioning}
\usetikzlibrary{backgrounds}
\usepackage{authblk}

\newtheorem{theorem}{Theorem}[section]

\newtheorem{lemma}[theorem]{Lemma}

\newenvironment{proof}[1][Proof]{\begin{trivlist}
\item[\hskip \labelsep {\bfseries #1}]}{\end{trivlist}}

\newcommand{\qed}{\nobreak \ifvmode \relax \else
      \ifdim\lastskip<1.5em \hskip-\lastskip
      \hskip1.5em plus0em minus0.5em \fi \nobreak
      \vrule height0.75em width0.5em depth0.25em\fi}

\usepackage{float}
\usepackage[toc,page]{appendix}
\usepackage[noadjust]{cite}

\usepackage[colorlinks,citecolor=blue]{hyperref}
\usepackage[inline]{enumitem}

\usepackage{mathtools}

\usepackage{enumitem}
\usepackage{algorithm}
\usepackage[noend]{algpseudocode}

\makeatletter
\def\BState{\State\hskip-\ALG@thistlm}
\makeatother

\graphicspath{ 
  {../Codes/Figures/}
}

\usepackage{amssymb}
\usepackage[square,sort,comma,numbers]{natbib}

\newcommand{\bitem}{\begin{itemize}}
\newcommand{\eitem}{\end{itemize}}
\newcommand{\benum}{\begin{enumerate}}
\newcommand{\eenum}{\end{enumerate}}
\newcommand{\beq}{\begin{equation}}
\newcommand{\eeq}{\end{equation}}
\newcommand{\beqs}{\begin{equation*}}
\newcommand{\eeqs}{\end{equation*}}

\title{Nonparametric Identification of Latent Dimension under Heavy-Tailed and Mixed-Signals}

\author{Chetkar Jha} 
\affil{University of Southampton Delhi}

\date{\today} 

\begin{document}

\bibliographystyle{cj}
\maketitle

\begin{abstract}
Principal component analysis and factor analysis are foundational to the study of high-dimensional data, yet their efficacy depends entirely on correctly identifying the latent dimension $r$. While parallel analysis (PA) is widely regarded as the gold standard for this task, its performance suffers in high-dimensional regimes consisting of mixed signals and/or heavy-tailed error distributions. This arises because under the preceding conditions, column-wise permutation mechanism (or any modern variations) applied to the data matrix $\textbf{X}$ fails to destroy the signal, leading to an inflated null spectrum. To address this, we propose a nonparametric solution. We show that the latent dimension is uniquely preserved in the degree of order within the rank-transformed data matrix, which restores spectral exchangeability under permutation. We prove the consistency of Rank PA under relaxed conditions, requiring neither the pervasiveness of factors nor restrictive fourth moment assumptions. We establish that Rank PA can consistently estimate the number of factors up to the maximum of $O(n^{1/3 - \tau})$ where $\tau >0$. The robust performance of our remedy is demonstrated through numerical simulations and an analysis of the FRED-MD dataset.
\end{abstract}
keywords{Factor models, Rank-based inference, Parallel analysis, High-dimensional data, Marchenko-Pastur law}

\section{Introduction}
 Principal Component Analysis (PCA) was proposed by Pearson (1901) \cite{pearson1901} and rediscovered by Hotelling (1933) \cite{hotelling1933} as a method to discover a set of (fewer) independent components that can explain a majority of variation in a set of correlated variables. In essence, the principal component model approximates the correlated variables as a weighted linear sum of (fewer) independent principal components. Independently, Spearman (1904) \cite{spearman1904} discovered that the $p$ test scores of $n$ students were correlated. This motivated him to propose a theory of intelligence using a single factor. Thurstone (1935) \cite{vom} expanded the scope of the single-factor model to multiple factors, leading to the present-day factor analysis (FA). In FA, the correlated variables can be written as a weighted linear sum of common factors and an error term in the normal linear factor model (Anderson (1958)\cite{anderson1958introduction}) with the weights as factor loadings
\be\label{factor.model}
\boldsymbol{X}_{n \times p} = \boldsymbol{\Lambda}_{n \times r} \boldsymbol{f}_{r \times p} + \mathcal{E}_{n \times p},
\ee
where $\boldsymbol{X}_{n \times p}$ is the underlying data, $\boldsymbol{\Lambda}_{n \times r}$ is the latent factor scores, $\boldsymbol{f}_{r \times p}$ is the factor loadings, $\mathcal{E}_{n \times p}$ is the error matrix, $n$ is the number of samples, $p$ is the number of features, and $r$ is the true number of factors.

\subsection{PCA and FA in the literature}
Researchers have extensively relied on PCA and FA for extracting the low-dimension representation of datasets for a simpler understanding of the underlying data. In particular, PCA and FA have been extensively used across several study areas such as psychology, psychiatry, medicine, spectroscopy, chemistry, biology, physiology, geriatrics, social science, population genetics, bio-informatics, economics, chromatography, see Jolliffe (2002) \cite{jolliffe2002}, Cattell (1988) \cite{cattellfactor1988}, Lopes (2014) \cite{lopes2014}. Jolliffe and Cadima (2016) \cite{jolliffe2016} describe the basic idea of PCA as a low-dimensional representation of large datasets in an adaptive and insightful way. They showed that PCA could be adapted for different datasets such as functional PCA being used for studying chemical spectroscopy or PCA being adapted for temporal atmospheric datasets. Notably, PCA has also been adapted for population genetics data, see Edwards and Cavalli-Sfoza (1963) \cite{edwards1963}. However, recently Elhaik (2022) \cite{elhaik2022} showed that PCA-based applications in population genetics can be manipulated to produce contradictory results making a case for developing robust PCA methods. However, with the pressing need to deal with large datasets in study areas such as image processing, machine learning have led to newer variants of PCA such as principal component pursuit problem and sparse PCA, see Huang et al. (2008)\cite{huang2008}, Witten et al. (2009) \cite{witten2009}, Journee et al. (2010) \cite{journee2010}, Cand\'{e}s et al. (2011)\cite{candes2011robust},  Huang and Wei (2019)\cite{huang2019}, Bertsimas and Kitane (2022) \cite{bertsimas2022}.

On the other hand, Cattell states that FA is a means of creating concepts and is not merely concerned with checking the fit (Cattell (1988)\cite{cattellfactor1988}). Costello and Osborne (2005)\cite{costello2005} detailed the recommended practice for implementing exploratory factor analysis. Lopes (2014) \cite{lopes2014} reviewed existing FA approaches. Lopes and West (2004) \cite{west2004} proposed a reversible jump Markov Chain Monte Carlo (RJMCMC) for the Bayesian factor analysis implementation where they also estimated the number of factors. Subsequently, many methods have been proposed for estimating sparse Bayesian factor analysis such as Carvalho et al. (2008) \cite{carvalho2008}, Knowles and Ghahramani (2011) \cite{knowles2011}. FA is also popular in econometrics literature as it breaks down the economic data in terms of latent factors see Bai and Ng (2002) \cite{bai2002determining}, Stock and Watson (2002) \cite{watson2002}, Onatski (2009) \cite{onatski2009}, Bai and Ng (2008)\cite{bai2008factor_review} (for a detailed review). However, the classical econometric literature assumes normal errors and pervasive factors which is typically not reflected in the economic data.
FA is also adapted in insightful ways for newer datasets. For instance, Wirth and Edwards (2016) \cite{wirth2016} adapted FA for item response data for ordered-categorical data, Bayesian FA (Zhou et al. (2023) \cite{zhou2023}) was applied for detecting genes that are affected by perturbations in CRISPR screening and FA was applied to study the practice of health literacy questionnaire (in Malay language) related to Covid-19 (see Dalawi and Aimran (2025) \cite{covidfactor2025}). Additionally, FA and PCA (PCA in particular) are increasingly being used as dimension reduction methods, see (Cekic et al.(2022)\cite{cekic2022}, Szarfraniec-Siluta et al.(2023)\cite{siluta2023}). 

\subsubsection{Estimating the Number of Factors and Components}
In theory, PCA and FA assume the true number of factors to be known. However, the practical implementation of PCA and FA hinges on knowing the true number of components and factors \emph{a priori}. Several authors such as Brown (2014)\cite{brown2014}, Decelle et al.(2011)\cite{decelle2011}, Onatski (2010)\cite{onatski2010factor}, Owen and Wang (2016)\cite{owen16} have stated that selecting the true number of components and/or factors is the primary challenge. 

\subsection{Related Work}
  Existing literature for estimating the number of components or factors can be categorized into three broad approaches, see Owen and Wang (2016) \cite{owen16} for a systematic review. The first approach relies on the scree plot - the line plot of ordered eigenvalues of the sample correlation matrix (Cattell (1966) \cite{cattell1966}). The scree method entails a visual inspection for an \emph{elbow} and the number of components to the left of the elbow is used as the estimate. While traditional approaches rely on the visual inspection, Onatski (2009) \cite{onatski2009} quantified the elbow in terms of the maximum ratio of consecutive eigenvalue gap statistics. Recently, Caro and P\~{e}na (2025) \cite{caro2025} proposed a new eigenvalue ratio criterion for estimating the number of factors in multivariate time series. 
  The second approach treats the problem as a model selection problem. These methods maximize an information criterion to estimate the number of components such as Bayesian information criterion proposed by Schwarz (1978) \cite{schwartz1978}, Bai and Ng (2002) \cite{bai2002factor}'s information criteria $IC_1, IC_2, IC3$. Conti et al. (2014) \cite{bfactor2014} proposed a Bayesian factor analysis for simultaneously estimating the factor loadings and the number of factors. Recently, Fr\"{u}hwirth-Schnatter et al. (2025) \cite{schnatter2025} proposed a Bayesian approach for estimating the parameters in FA including the estimated number of factors in sparse (normal) latent factor model. Li et al. (2017) \cite{LI201776} modified Bai and Ng (2002) \cite{bai2002factor}'s information criteria for estimating the number of factors. Recently, Haslbeck and Bork (2024) \cite{haslbeck2024} proposed a new method for estimating the number of factors using out-of-sample prediction errors.
  The third approach estimates the number of components by comparing eigenvalues of the sample covariance matrix against an empirical threshold derived under a null model. Earlier works, such as Guttman (1954) \cite{guttman1954factor} and Kaiser (1960) \cite{kaiser1960} (Kaiser's criterion), estimated the number of factors as the number of eigenvalues of the sample correlation matrix that are larger than one. However, they fail to account for the sampling variability. The shortcoming of Kaiser's criterion motivated Horn (1965) \cite{horn1965} to propose a data dependent method called parallel analysis for empirically estimating the threshold. Later many improved versions of parallel analysis were proposed. Moreover, a select number of random matrix theory (RMT) inspired methods estimates the empirical threshold using RMT based approaches. For instance, Fan, Guo, and Zheng (2020) \cite{fan2019factor} improved upon the Kaiser's criterion by adjusting the threshold of the sample correlation matrix. Many fast and newer versions of parallel analysis such as deterministic parallel analysis (DPA), deflated deterministic parallel analysis (DDPA) (Dobriban and Owen (2019)\cite{dobriban2018dpa}), sign flip method (Hong and Dobriban (2020) \cite{randomsignflips}) also estimate the threshold using the RMT theory. Ke, Ma, and Lin (2020)\cite{ke2020ci} and Jha and Barnett (2023) \cite{jha2023} also proposed confidence intervals for the number of components in FA and PCA. Despite their sophistication, the current state of the art methods rely on the spectral properties and the asymptotic convergence of the sample correlation matrix to the identity in the high-dimensional set up. As we demonstrate, this makes them susceptible to mixed signals and heavy tailed error distributions conditions typically found in financial data.

%
\subsection{Parallel Analysis}
Horn (1965) \cite{horn1965} proposed parallel analysis as a data dependent approach to estimate the number of components. Parallel analysis uses the original data to generate multiple realizations from the null distribution. A popular version of parallel analysis (PA), proposed by Buja and Eyuboglu (1992) \cite{buja1992}, implements this scheme by permuting the column entries of the data matrix $\textbf{X}_{n \times p}$. Permuting the column entries destroys any signal of the data matrix. The permuted data matrix can be imagined as a realization of the noise matrix under the null model. The advantage of PA is that we can repeatedly permute the data matrix to generate multiple independent realizations of the noise matrix under the null model. Using the permuted data matrix, we can construct the empirical distribution of the singular values under the null. The percentiles of the empirical distribution of singular values can then be compared against the singular values of the data matrix to test whether the original data matrix has any signal components. Repeatedly removing the signal components from the data and then comparing the singular values of the data matrix against the ($50th$ or $95th$) percentiles of the empirical distribution of the singular values under the null can be used for estimating the number of (signal) components (or the number of factors) of the original data matrix, see Dobriban and Owen (2019)\cite{dobriban2018dpa}, Dobriban (2020) \cite{dobriban2017permutation}. Intuitively, the above procedure makes sense. However, until Dobriban (2020) \cite{dobriban2017permutation}, we lacked the mathematical understanding of the parallel analysis. 

 Recently, Dobriban (2020)\cite{dobriban2017permutation} proved the consistency of PA under certain (restrictive) conditions. Surprisingly, they remarked (in the Section 3.1 of their paper) that \emph{PA selects the number of perceptible factors as long as the signal strength is not too large}. A similar observation is made by Owen and Wang (2016) \cite{owen16} about having weaker signals. The rationale is that PA relies on destroying the signals by permuting column entries to create multiple realizations of the noise matrices. Particularly, in the existing versions of PA, when one or many signals have large signal strength then destroying the signal (one-by-one) will create realizations of the noise matrices with larger variance. It means that the realized singular values under the null distribution will be inflated and forcing us to overlook the relatively weaker signals. Dobriban and Owen (2019)\cite{dobriban2018dpa} called this phenomenon as a \emph{shadowing effect} of large signals. They proposed a deflated deterministic parallel analysis (DDPA) for tackling the shadowing effect. However, the faster and improved versions of PA such as DPA, DDPA, sign flip method still gets affected by the shadowing effect. In particular, the existing versions of parallel analysis are sensitive to large signals. Moreover, in the Section 2.1 of the Dobriban and Owen (2020)\cite{dobriban2017permutation} note that the PA can only estimate the number of factors when the signal strength or size is above a threshold which is not easily defined. Additionally, they provide the mathematical guarantee for PA under \emph{specific signal models} in Section 3 of their paper. 

\subsubsection{Limitations}
  Dobriban and Owen (2019) \cite{dobriban2018dpa}'s work gives an impression that parallel analysis and their variants are theoretically sound until it does not. Their argument require that the correlation of the data matrix $\textbf{X}_{n \times p}$ generated under the factor model in (\ref{factor.model}) converges to the identity matrix $\textbf{I}_{p \times p}$. However, this ignores the \emph{so-called} pathological case when the errors in (\ref{factor.model}) are heavy-tailed distributions and/or when the factor loadings are mixed (i.e. consisting of strong and weak factors). Unfortunately, the heavy-tailed distributions or mixture of factor loadings are quite common-place in econometrics data and hardly pathological conditions. Additionally, this is not limited to PA ( and their modern variations), in fact any method such as Fan et al. (2020)'s \cite{fan2020factor} that rely on the asymptotic convergence to the identity matrix also breaks down under these conditions. The real question is why? Consider

  \ba
  \frac{1}{n} \textbf{X} \textbf{X}^\top = \frac{1}{n}\Lambda \Lambda^\top + \frac{1}{n}\mathcal{E}. \mathcal{E}^\top 
  \ea

  Now, when the factor loadings are similar then it is true that under random permutations the signal structure is destroyed. In particular, the average over the random permutations would converge to a constant matrix, i.e, $\frac{1}{n}\Lambda \Lambda^\top \to \mathcal{C}$ where $\mathcal{C}$ is a constant matrix. Since the correlation matrix is invariant to the centering then the sample correlation matrix converges to the identity matrix when the errors in (\ref{factor.model}) are normally distributed. This asymptotical miracle falls apart under the two scenarios when the factor loadings are mixed (weak and strong) and/or when the errors are heavy-tailed distributions.

  When the factor loadings are mixed (weak and strong signal), then under the random permutations the signal contribution to the covariance matrix, $\frac{1}{n} \Lambda \Lambda^\top$ does not converge to a constant matrix. In particular, \emph{the signals are not destroyed under the random permutations} and the random permutations only smoothens out the final limit of the $\frac{1}{n} \Lambda \Lambda^\top$. This forces the sample correlation matrix to deviate away from the identity matrix even under the Gaussian errors. This is exactly the reason that Dobriban (2020) \cite{dobriban2017permutation} only cites the consistency result of parallel analysis under restrictive signal conditions. In fact, it is correct to say that the parallel analysis (in the current form, i.e., PA) would not be able to recover true number of latent dimension when the signals are a mixture of weak and strong factors. The similar observation holds for other methods that rely on the convergence of the correlation matrix to the identity matrix such as Fan et al. (2020) \cite{fan2020factor}'s ACT method. When the error distributions are heavy-tailed distributions even with bounded fourth moment conditions on the data matrix, $\frac{1}{n}\mathcal{E}\mathcal{E}^\top$ might be slow to converge to the identity matrix, i.e., 
  $$||\frac{1}{n}\mathcal{E}\mathcal{E}^\top - \textbf{I}_{p \times p}|| = O_p(n^{-\alpha}),$$
  where $\alpha$ is small. Therefore, for finite $n$, all the methods that rely on the asymptotic convergence of the sample correlation matrix converging to the identity matrix will be impacted. 
  This is demonstrated in the section \ref{s.numerical} where we consider errors to come from the student's t distribution with degrees of freedom to be $5$ and/or factor loadings to be a mixture of weak and strong signals. 
  
\begin{figure}
      \centering
      \includegraphics[width=\linewidth]{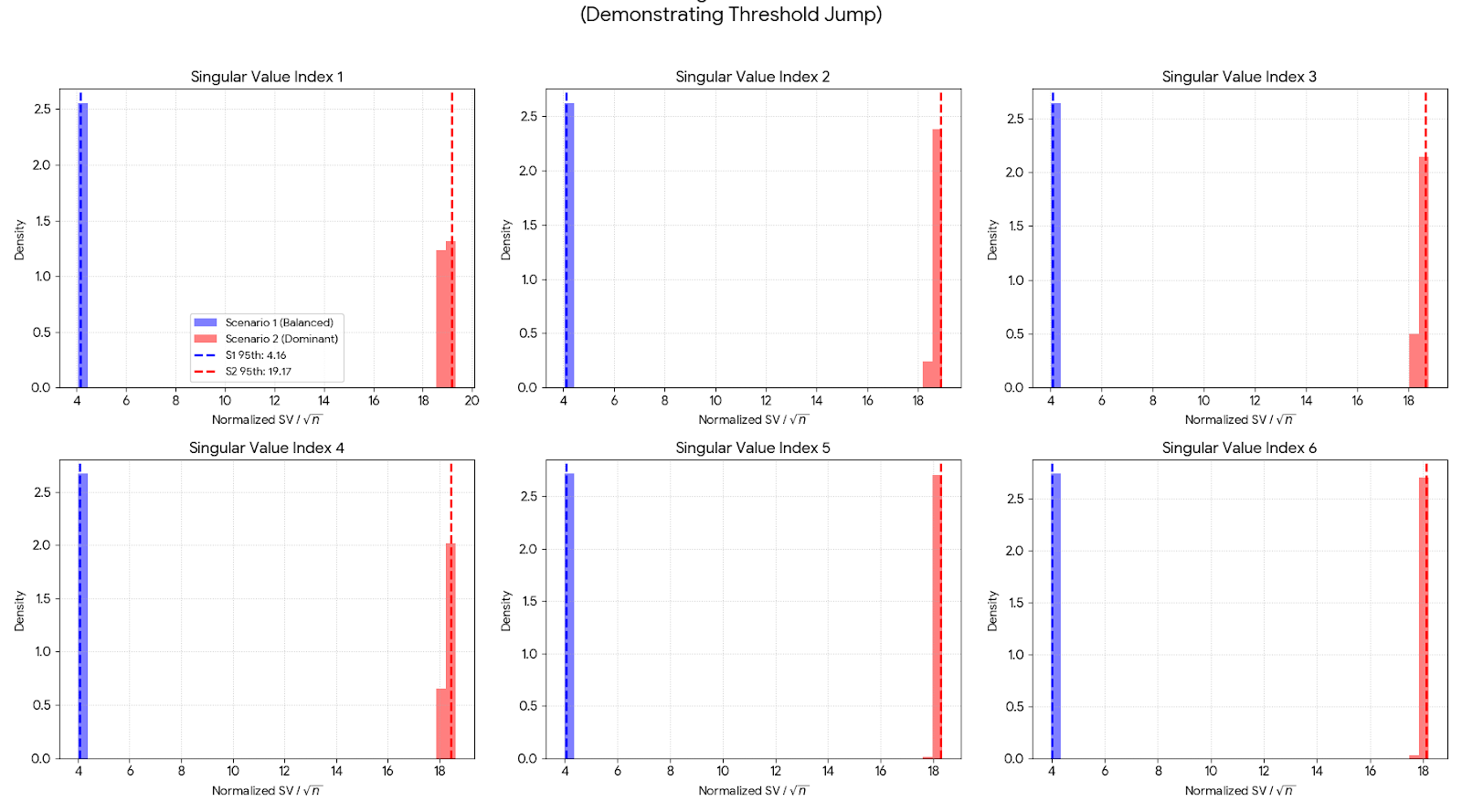}
      \caption{Histograms of top six singular values of the permuted data matrix when $(n,p) = (2000,200)$ and $K =12$. The blue histogram corresponds to the balanced case where every factor loading was $3$ while the red histogram correspond to the mixed factor loading when one loading values as $18$ and the rest of them were $3$.}\label{F:2}
  \end{figure}

\begin{figure}
      \centering
      \includegraphics[width=\linewidth]{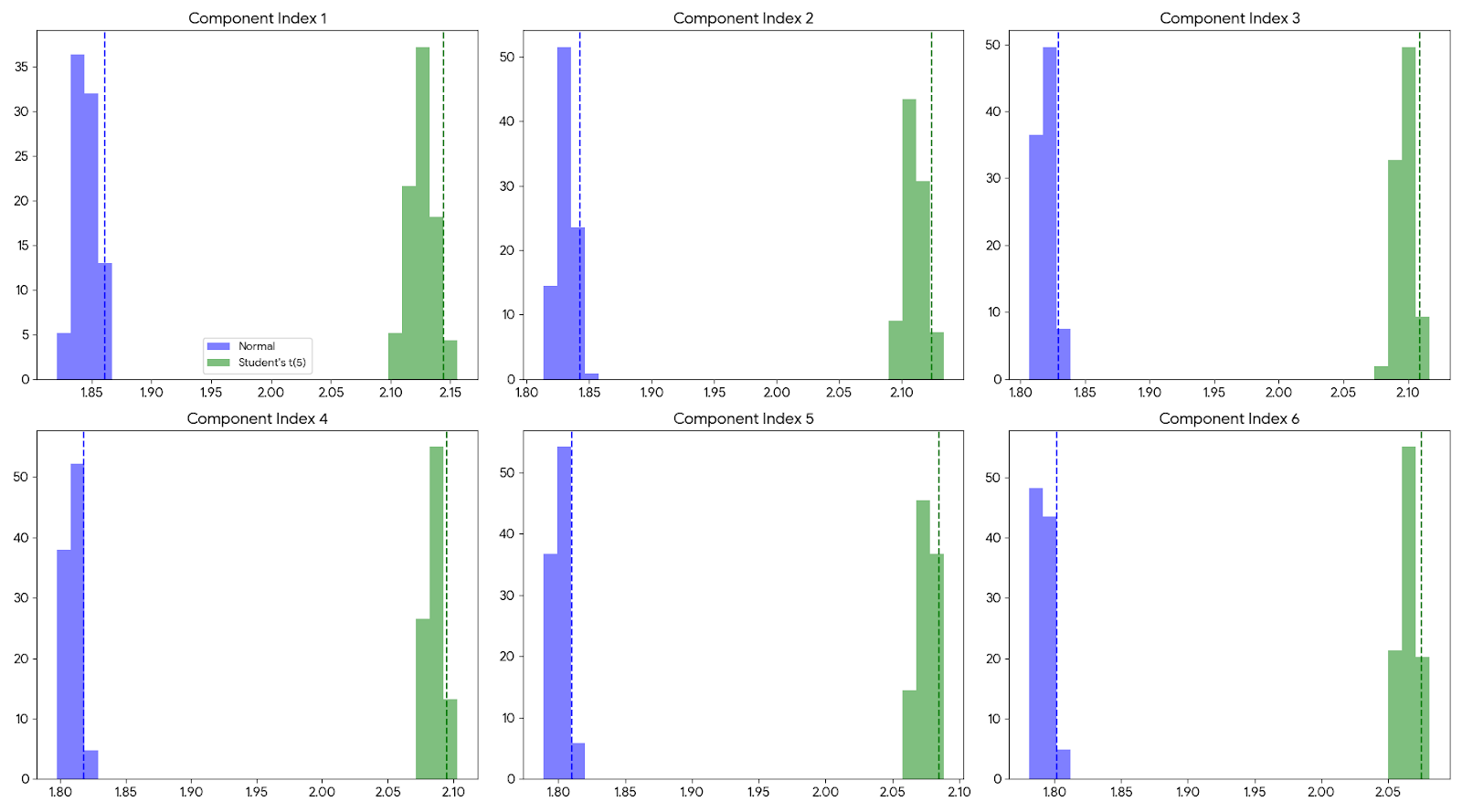}
      \caption{Histograms of top six singular values of the permuted data matrix when $(n,p) = (2000,200)$ and $K =12$. The blue histogram corresponds to the normal case where every factor loading was $1$ while the red histogram correspond to the student's t (5).}\label{F:3}
  \end{figure}

  The above phenomena is clearly depicted in the figures \ref{F:2}-\ref{F:3} where we see the histograms of top six singular values of the permuted data matrix $\frac{1}{\sqrt{n}}\textbf{X}_{\pi}$ shifts (to the right) when the loading factor changes from the balanced loadings to a mixture of loadings with one dominant factor (figures \ref{F:1}-\ref{F:2}) and when the factor loading is small but the error distribution changes from the normal errors to student's t with $5$ degrees of freedom.  The above shift inflates the null value threshold and negatively affecting the performance of PA, their modern variants and other methods relying on the asymptotic convergence of the sample correlation matrix to the identity matrix.
    Recall that the genesis of the parallel analysis and other methods was to account for the sampling fluctuations which was not considered by Kaiser (1960)'s criterion \cite{kaiser1960}. However, under the above scenario of a  mixture of weak and strong factors and heavy tailed error distributions with weak factors, the parallel analysis and the other methods such as Fan et al. (2020) \cite{fan2020factor} fail to do just that.


 \subsubsection{Intuition}
 We postulate that the signals in the data matrix correspond to the orderly arrangement of the columns of the data matrix and mere column-wise (random) permuting of the data matrix might fail to fully destroy the signal ( see figures \ref{F:1}-\ref{F:3}). The argument is nuanced. In particular, we show that the column-wise (random) permutations of the rank data matrix destroys the signal which therefore can be used to generate an appropriate null distribution. We use this observation to propose a rank-based nonparametric version of parallel analysis called Rank PA. Our approach demands that we consider the rank data matrix for estimating the number of components. Rank data matrix is of the same dimension as the data matrix with every $(i,j)^{th}$ entry in the rank data matrix correspond to the rank of $(i,j)^{th}$ entry in the $j^{th}$ column of the original data matrix $\textbf{X}$. As in PA, we permute (column-wise) the rank data matrix to generate multiple realizations of the noisy rank matrices. Then, we compare the singular values of the rank matrix of the original data matrix against the percentiles of the empirical distribution of rank singular values under the null for estimating this \emph{degree of order} of the rank data matrix. In fact, we discover that this degree of order is the number of signal components in the data matrix. Later in the paper, we compare our approach with the existing versions of PA.



Although rank-based methods are widely considered in many areas of statistics. However, to the best of our knowledge, the rank-based PA has not been considered for estimating the number of components of FA or PCA. Our contribution is as follows: i) we proposed a new rank-based PA for estimating the number of components of FA or PCA, ii) we prove that our method is consistent under less restrictive conditions when the signal strengths is large and/or the number of components to be $O(n^{1/3})$. The rest of the paper is structured as follows. In Section \ref{s.main}, we introduce the notations, the underlying factor model, and our method. In Section \ref{s.main}, we also propose our method and establish theoretical properties. In Section \ref{s.numerical}, we use the numerical simulations to establish the superior and robust performance of our method. In Section \ref{s.data}, we apply our method to the real data analysis.
In Section \ref{s.discussion}, we close out the paper with a discussion. The mathematical proof is included in the Supplement. 

\section{Main Formulation}\label{s.main}
\subsection{Notations}\label{s.notation}
Consider $\textbf{X} =\left(x_{ij}\right)_{i=1,\cdots,n, j=1,\cdots,p}$ a data matrix consisting of continuous entries with dimension $n \times p$. Let $\textbf{R} =\left(r_{ij}\right)_{i=1,\cdots,n, j=1, \cdots,p}$ denote the corresponding (column) rank matrix with each entry $r_{ij}$ denoting the rank of $x_{ij}$ in the $j^{th}$ column of $\textbf{X}$. Let $\pi =(\pi_1, \cdots, \pi_p)$ denote a permutation array permuting the column entries of a matrix independently, i.e., $\textbf{X}_{\pi}(i,j) = \textbf{X}(\pi_j(i), j)$ for any $1 \le j \le p$. Let the singular values of $\textbf{X}$ be arranged in descending order, i.e., $\lambda_1 \ge \lambda_2 \ge \cdots \ge \lambda_p$. Similarly, let the singular values of $\textbf{R}$ be arranged in descending order, i.e., $\lambda^R_1 \ge \lambda^R_2 \ge \cdots \ge \lambda^R_p$. (In the supplement, we use $\mu_i()$ to denote the eigenvalue of a square matrix).
We use the notation that $\textbf{A}_{\pi} \stackrel{d}{=} \textbf{A}$ to denote the entries of $\textbf{A}_{\pi}$ obtained after column-wise (random) permutations $\pi$ have the same distribution as the entries of $\textbf{A}$, where $\textbf{A}$ is any real data matrix. Here in this paper, we always assume that the permutation array is random, i.e, each column level permutation is achieved by uniformly sampling (without replacement) over $\{ 1, \cdots, n\}$ and every column level permutation is independent of other columns. Let $\sigma$ denote the (common) standard deviation of $\textbf{R}$ which is $\sqrt{\frac{n^2-1}{12}}$, then we denote the scaled rank matrix $\textbf{R}^\star$ as $\textbf{R}/\sigma$.
We use $||\cdot ||_{op}$ and $||\cdot||_F$ to denote the operator norm and Frobenius norm, respectively. 
Without loss of generality, we assume the data matrix $\textbf{X}$ is centered.
We assume that the observations are generated from the standard factor model, where $x_i = (x_{i1}, \cdots, x_{ip})^{T}$.
The standard factor model is given as
\be\label{std.factor}
x_i = \Lambda \eta_i + \epsilon_i,
\ee
where $\Lambda$ is the $p \times r$ factor loading matrix with factor loadings ${l_{jk}}$, where $j=1, \cdots, p$, $k=1, \cdots, r$, $\eta_i$ is the vector of length $r$ consisting of factor values, and $\epsilon_i = (\epsilon_{i1}, \cdots \epsilon_{ip})^T$ is the error vector of length $p$. The standard factor model under the normal errors would be referred as the standard normal factor model otherwise we do not make explicit assumptions about the error distribution.
The standard factor model in (\ref{std.factor}) can be written in the matrix terms as follows

\be\label{factor.model}
\textbf{X} = \textbf{F}\Lambda^\top  + \mathcal{E},
\ee
where $\textbf{F}$ is the matrix of dimension $n \times r$ and $\mathcal{E}$ consisting of factor values and residual values, respectively. The above factor model in (\ref{factor.model}) is not identifiable and we require restriction that $\Lambda \Lambda^\top = \textbf{I}_{r\times r}$ to make the model identifiable, see Bai and Ng (2002) \cite{bai2002factor}. Before, we introduce rank PA, we prove that PA can lead to inconsistent estimator. For this, we introduce the spiked covariance model.

\subsection{Spike Covariance Model}
For simplifying the study of covariance matrix under the high-dimensional set up, Johsntone (2001) \cite{johnstone2001distribution} proposed a spiked covariance model as a diagonal matrix with few eigenvalues larger than one and the rest of them being one. The spiked covariance model is given as
\be\label{spike}
\boldsymbol{\Sigma} = \sum_{j=1}^K h_j u_j u_j^\top + \textbf{I}.
\ee

We use the spiked covariance model set up to study the performance of PA under mixed factor loadings and non-pervasive factors,i.e., when $j^{th}$ factor loading is shared across $p_j$ many factors with $p_j$  is smaller than $p$.
\begin{theorem}{(PA under non-pervasive and mixed factor loadings)}
Consider the data matrix $\mathbf{X}$ generated under (\ref{factor.model}) and satisfying the spiked covariance structure in (\ref{spike}). From the covariance model, we see that 
$$\mu_j = \begin{cases}
 h_j + 1 ,& j \le K\\
 1 ,& j > K\\
\end{cases}
$$
Let $l_1$ be factor loading of the dominant factor and $l_k$ be the $k^{th}$ factor loading of a weak factor with $p/n \to \gamma$ as $n \to \infty$. Then when the $k^{th}$ factor loading is non-pervasive, i.e., $k^{th}$ factor loading is shared among $p_k$ number of columns, 
the PA estimator $\hat{K}$ is inconsistent and fails to recover the $k^{th}$ factor when
$l_1^2 > l_k^2 p_k$
\ba
P(\hat{K} < k) \to 1,
\ea
where $k \le K$ the true number of factors.
\end{theorem}

\begin{proof}
Under the null obtained by permuting the column entries of $\textbf{X}$, we get the spectral number of $|| \textbf{X}_{\pi}||_{op}$ convergence is governed by the BBP phase transition (Baik et al. (2005)\cite{baik2005phase}), i.e., 
\be
\frac{1}{\sqrt{n}}||\textbf{X}_{\pi}||_{op} \stackrel{p}{\to} \begin{cases}
    1 + \sqrt{\gamma} ,& l_1 < 1 + \sqrt{\gamma} \\
    \sqrt{l_1^2 (1 + \frac{\gamma^2}{l_1^2})}, & l_1 > (1 + \sqrt{\gamma})
\end{cases}
\ee

However, under the assumption of non-pervasive factors, i.e., when the $k^{th}$ factor loading is shared across $p_k$ columns. Under the spiked model (\ref{spike}), the $k^{th}$ sample singular value converges to the BBP limit governed by $\mu_k$. For pervasive/non-pervasive factors, $\mu_k = l_k^2 p_k + 1$. Then, 
\ba
\lambda_k(\frac{1}{\sqrt{n}} \textbf{X}) = \sqrt{l_k^2 p_k + 1}.
\ea

Therefore, for PA to select the $k^{th}$ factor with probability one when there is a dominant factor and  a non-pervasive weak factor is to satisfy $|| \frac{1}{\sqrt{n}}\textbf{X}_{\pi}||_{op} <= \lambda_k(\frac{1}{\sqrt{n}} \textbf{X})$. However, when this condition is not met (as illustrated in the main paper), i.e. , $P(\hat{K} < k) \stackrel{p}{\to} 1$.
\end{proof}

Note that the above breakdown of PA is the failure of permutation based methods to destruct the signal in presence of a dominant and pervasive weak factors. The implication of this breakdown extends to the many variants of PA and the other methods which rely on the homogeneity of factor loadings. Then one may argue that this is a pathological case but it is a common-place situation in the econometric data particularly in finance. Moreover, the birth of parallel analysis and alternative methods for estimating the number of factors was to address the shortcoming of Guttman (1954) \cite{guttman1954factor} and Kaiser criterion (1960) \cite{kaiser1960}.

\subsection{A General Signal Model}
Consider a general model where the data matrix can be written in terms of signal and noise components. Let the signal matrix $\textbf{S}$ be a matrix of dimensions $n \times p$ and the noise matrix $\textbf{N}$ be a matrix of dimension $n \times p$. Then, we can write
\be\label{Signal}
\textbf{X} = \textbf{S} + \textbf{N}.
\ee

Without loss of generality, we assume that \textbf{X} is centered. It is easy to see that centering the data matrix or the signal will not have any effect on the number of signal components. First, we consider the special case of the signal model in (\ref{Signal}) where the variance of the diagonal entries of the noise matrix is one. We argue that this is not a special case because this can be achieved by pre-multiplying the equation in (\ref{Signal})
by a diagonal matrix with the $i^{th}$ diagonal entry is one over the standard deviation of the $i^{th}$ error term in (\ref{std.factor}) provided it is greater than zero. (If a particular column has a zero standard deviation then that is of no interest in the PCA or FA and we discard it). In the general case, we rename the data matrix as \textbf{X} and the signal strength of \textbf{S} becomes the signal to noise ratio (SNR), i.e., the minimum signal strength to the variance of noise entries. The rest of the analysis is carried out under the above general case. Moreover, we require the noise matrix to satisfy some additional conditions. In particular, it is assumed that for the standard factor model with the maximum possible number of factors the errors are uncorrelated,i.e., the fluctuations under the full standard factor model are not systematic as all the factors are included in the model.

\begin{enumerate}[label = A\arabic*]
\item\label{a1} (Unit Variance): The variance of the error entries in (\ref{std.factor}) is one,
\item\label{a2} (Uncorrelated) : The error terms in (\ref{std.factor}) are uncorrelated and the factors and the error terms in (\ref{std.factor}) are also uncorrelated
\item\label{a3} (Invariance): The noise matrix should be invariant under any column wise permutation $\pi$, i.e., $\textbf{N}_{\pi} \stackrel{d}{=} \textbf{N}$.
\end{enumerate}

In PA, the column entries of the data matrix are permuted to destruct the signal present in the data matrix, which is subsequently used for estimating the number of components. It is easy to observe that random permutation of the column entries of the data matrix $\textbf{X}$ destroys both the signal and the ordering structure of the column entries of the data matrix. In this context, we consider the column-wise rank matrix \textbf{R} whose $(i,j)^{th}$ element is the rank of \textbf{X}(i,j)
among all the $n$ entries of the $j^{th}$ column of $\textbf{X}$. Note that \textbf{R} is determined by the order of the column entries of $\textbf{X}$. Then, it is natural to ask how random permutation of the column entries of $\textbf{X}$ affects the rank data matrix $\textbf{R}$.

\subsection{Equivalence Results on Rank Data Matrix}

\begin{lemma}[Rank-Signal Equivalence]\label{rank.signal}
For the signal model in (\ref{Signal}), we define $\textbf{R}$ as the column-wise rank matrix of $\textbf{X}$. Additionally, we assume that the noise distribution in (\ref{Signal}) is absolutely continuous with respect to the Lebesgue measure. Then, we show the following
\begin{enumerate}[label=C\arabic*.]
\item When $||\textbf{S}||_{F} =0$, then $\textbf{R}_{\pi} \stackrel{d}{=} \textbf{R}$,
\item When $||\textbf{S}||_{F} \neq 0$, then $\textbf{R}_{\pi} \stackrel{d}{\neq} \textbf{R}$.
\end{enumerate}
\end{lemma}

\begin{proof}
The proof is given in the Appendix.
\end{proof}

A direct consequence of the lemma \ref{rank.signal}
is that we establish an equivalence between the column-wise non-invariance (invariance) of $\textbf{R}$ under random column-wise permutation with the presence (absence) of the signal. In simpler words, if the signal is destroyed by randomly permuting the column entries of the data matrix $\textbf{X}$ then the column-entries of \textbf{R} is going to be invariant. Note that the rank data matrix $\textbf{R}$ does not admit a decomposition like in (\ref{Signal}). However, the above lemma establishes a necessary and conceptual equivalence between the signal matrix $\textbf{S}$ and the rank data matrix $\textbf{R}$.  In particular, we will use the equivalence to establish theoretical guarantees for rank PA. An important facet of the lemma \ref{rank.signal} is that we do not have any conditions on either the signal or signal strength as is the case with the existing theory for PA in Dobriban (2020)\cite{dobriban2017permutation}. 
It is with this foresight that we had remarked that the latent dimension is a feature of the ordered property of the rank matrix. This will become abundantly clear later. For expanding the scope of our approach to continuous non-normal error distributions, we require the columns of the noise matrix to be independent.

\begin{lemma}[Column Ranks Independence]\label{rank.ind}
For the signal model in (\ref{Signal}), we define $\textbf{R}$ as the column-wise rank matrix of $\textbf{X}$ where the entries of $\textbf{X}$ are assumed to be continuous. Then, we show that when $||\textbf{S}||_F=0$ and the columns of $\textbf{N}$ are independent, then the columns of $\textbf{R}$ are going to be independent.
\end{lemma}

\begin{proof}
The proof is in the Appendix.
\end{proof}

In general, the errors in equation (\ref{std.factor}) are assumed to be normal then the assumption \ref{a2} automatically implies independence of error column vectors. However, for datasets with
continuous non-normal errors (for instance, the macroeconomic panel data of Fred-MD) we require the error columns to be independent for theoretical validity. This helps us expand the scope of our approach to include continuous non-normal error distributions. Lemma \ref{rank.signal} establishes the crucial link between the presence/absence of signal with the column-invariance of the rank data matrix. 

We now show that under the null, the $i^{th}$ singular values of $n^{-1/2} \textbf{R}^\star$ is bounded and the Monte Carlo estimation using the independent permutations converges to $\lambda_i(n^{-1/2} \textbf{R}^\star)$.

\begin{theorem}[Asymptotic Consistency]\label{asym} 
For the continuous and centered data matrix $\textbf{X}$ under the general signal model (equation (\ref{Signal})) with $p/n \to \gamma \in (0, \infty)$ satisfying the following conditions
\begin{enumerate}
\item Null-Model level assumptions
\begin{enumerate}
\item \emph{No Signal}. Under the null, we have $||\textbf{S}||_F \to 0$.
\item \emph{Noise Invariance}. The distribution of the noise is invariant under permutations $\textbf{N}_{\pi} \stackrel{d}{=} \textbf{N}$ with the equality in distribution is taken under joint distribution and the permutation $\pi$ is randomly chosen.
\end{enumerate}
\item Error level assumptions \begin{enumerate}
\item \emph{Uncorrelated Normal Errors or Column-wise Independent Noise Columns}. 
\item \emph{Errors and Factors are uncorrelated}
\item \emph{Errors have unit variance and they are absolutely continuous with respect to the Lebesgue measure}
\end{enumerate}
We show that the Monte Carlo estimation of $j^{th}$ singular values of $\textbf{R}$ using independent random permutations is consistent, i.e.
\ba
 \frac{1}{k} \sum_{k=1}^K\lambda_j(n^{-1/2} \textbf{R}_{\pi_k}/\sigma) \to \lambda_j(n^{-1/2} \textbf{R}/\sigma).
 \ea
\end{enumerate}
\end{theorem}

\begin{proof}
Please see the Appendix.
\end{proof}

The above theorem gives the convergence result for the Monte Carlo estimate under the null. These conditions include null-model level assumptions (i.e. no signal in equation (\ref{Signal})) and the error-related assumptions of the noise matrix. Under the null, when there is no signal, we have $||\textbf{S}||_F$ going to zero and the column-wise permutations of the noise matrix is invariant. These assumptions are typical. The error-related assumptions are the standard assumptions (see (\ref{a1})-(\ref{a3})) in the factor model with an exception that we require column-wise independent noise matrix. 
The proof relies on showing that under the null, the $j^{th}$ singular values are bounded and the $j^{th}$ eigenvalues is a Lipschitz function. Notably, we do not require the fourth moment bound on the data matrix $\textbf{X}$ as the fourth moment of $\textbf{R}^\star$ is automatically bounded by the construction. This is a crucial edge of our method against existing approaches as they require the fourth moment of $\textbf{X}$ to be bounded (Bai and Ng(2002)\cite{bai2002determining}, Dobriban and Owen (2017)\cite{dobriban2017}, Dobriban (2020)'s PA \cite{dobriban2017}, Fan et al. (2019) ACT \cite{fan2008factor}) which considerably limits their real-world applicability.
Furthermore, for generalizing the scope of our approach to continuous non-normal distributions we assume the column-wise independence of the noise matrix whereas the other competing methods are not suitable for continuous heavy-tail distributions.

\subsection{Rank PA}
Motivated by our understanding of signal data matrix and theorem \ref{asym}, we propose an elegant method for estimating the number of factors in the factor model or the number of components in the principal component model.
This utilizes the idea that column-wise (random) permuting destroys the order of the data matrix and consequently destroys the order of the rank data matrix. Using this insight, we propose a Rank PA in the algorithm \ref{alg:2}.

\begin{algorithm}
\caption{Rank PA}\label{alg:2}
\begin{algorithmic}[1]
\State Input the data matrix $\textbf{X}$ of dimension $n \times p$. 
\State Compute the column-wise rank matrix of $\textbf{X}$ as $\textbf{R}$ with entries of every column in $R$ varying over $\{1, \cdots, n\}$.
\BState Let the singular values of the Singular Value Decomposition (S.V.D.) of $\textbf{R} = \textbf{U} \textbf{D} \textbf{V}^\top$  be stored as $\{ \lambda^r_i\}_{i=1}^p$.
\BState Pick a large integer value of $M$ and permute the column entries of $\textbf{R}$ to create $M$ permuted versions of column rank matrix $R$. Store it in $\{ \textbf{R}_{\pi
_j}\}_{j=1}^M$ with $\pi_j$ denote the permutation operator.
\BState Perform the S.V.D. for $\{\textbf{R}_{\pi_j} \}_{j=1}^M$ and compute the $95^{th}$ percentile of the popular singular values as $\{\lambda^{R^0}_i \}_{i=1}^p $.
\BState Set $i =1$
\While{ $\lambda^R_i > \lambda^{R^0}_{i, .95}$ }{

  $i = i + 1$
 }
 \EndWhile
\BState Output $\hat{r} = i - 1$.
\end{algorithmic}
\end{algorithm}

Technically, in the Step $7$ of algorithm \ref{alg:2} should be $\lambda^R_i/ \sigma > \lambda^{R_0}_{i, 0.95}/\sigma$, where $\sigma = \sqrt{\frac{n^2-1}{12}}$. However, $\sigma$ is constant and greater than zero therefore canceling out from both sides.
Notice the algorithm \ref{alg:2} is deceptively similar to PA proposed by Buja and Eyboglu (1992)\cite{buja1992}. A simple observation suggests that shadowing effect would be muted issue with rank PA when compared to PA and its variants. This is  because the rank data matrix does not depend on the signal strength. It has more to do with the ordering of the column entries of $\textbf{X}$.

\begin{theorem}[Rank PA Consistency]\label{main.thm}
Assume that the data matrix $\textbf{X}$ is generated from (\ref{factor.model}). Denote the corresponding column-wise rank matrix for the data matrix $\textbf{X}$ as $\textbf{R}$. Moreover, assume that null-model level assumptions and error assumptions in the theorem \ref{asym} are satisfied. Then our method can estimate the maximum number of factors $r$ up to the maximum order of $O(n^{1/3 - \tau})$ for $\tau >0$, i.e., the estimate proposed in the algorithm \ref{alg:2} is consistent.
\ba
P(\hat{r}= r) \to 1, \text{ as n} \to \infty.
\ea
\end{theorem}

\begin{proof}
    Please see the appendix for the proof.
\end{proof}

It is also important to note that our method does not ostensibly require any restrictive moment condition on the signal or the delocalization of the factors which was the theoretical foundation proposed by Dobriban (2020)\cite{dobriban2017permutation}. These conditions are restrictive as they do not offer the theoretical guarantee for PA in a more general non-parametric set up which has made PA popular. In contrast, our results extends to the continuous non-normal errors as long as the error columns are independent. Another important aspect of our approach is that we can estimate the growing number of factors or components up to the maximum of $O(n^{1/3})$. We believe that Rank PA is truly a non-parametric extension of PA that is applicable under general conditions. This is demonstrated through numerical comparisons.

\section{Numerical Comparisons}\label{s.numerical}
We compare the performance of our method against other competing methods on several simulation scenarios. The scenarios are of two types: i) When the loading factors are the same for all the factors, ii) When the loading factors are a mixture. Moreover, we vary the model type over Principal Component Model, Factor Model, the type of the error distributions over the normal distribution and Student's t distribution with $5$ degrees of freedom, the true number of factors and the loading factor. The loading factor is a measure of the signal strength. Additionally, we study the numerical simulations where $r$ is of the order $O(n^{1/3})$ or larger. Here, we vary the number of factors $r$ over $\{12, 14, 18\}$ for the data matrix with $n = 2000$ as $12$ is  $\lfloor 2000^{1/3} \rfloor$ where $\lfloor \cdot \rfloor$ is the greatest integer function. Moreover, we varied $p$ over $\{200, 400\}$. For the first scenario, we compiled the accuracy rate of our method and the competing method in the tables \ref{T:S1}-\ref{T:S4}. Tables \ref{T:S1}-\ref{T:S2} gives the accuracy comparison for the low signal strength scenario and the high signal strength scenarios as the loading factor is varied for $p=200$. The low loading scenario results in zero accuracy rate for Bai and Ng's (2002) \cite{bai2002determining} $IC_1, IC_2, IC_3$, demonstrating that their method is sensitive to the signal strength. DPA and Fan's ACT method do better than Bai and Ng (2002)'s method. However, DPA, Fan's ACT, and PA do poorly when the underlying error distribution is from the Student's t distribution, suggesting that they are sensitive to distributional level assumptions. In table \ref{T:S2} when the loading factor is high, then both PA and Fan's ACT method have far better performance, suggesting that the two methods are sensitive to the signal strength. However, DPA's performance is bad in both cases. In tables \ref{T:S3}-\ref{T:S4} we see that the DPA's performance improves when the aspect ratio changes from $0.2$ to $0.4$. Similar improvement is noticed in the performance of other methods as well. Based on the tables \ref{T:S1}-\ref{T:S4} we can easily see that Rank PA is far less sensitive to the distribution, the number of factors, error distribution, and the aspect ratio. Clearly, Rank PA has the superior performance among all methods in the tables \ref{T:S1} and \ref{T:S3}. Interestingly, Rank PA does well for $r > O(n^{1/3})$.

\begin{table}
\centering
\caption{Low Signal Mixture Scenario: The accuracy comparison between different methods for estimating the number of components in PCA and FA when the data matrix was of the size $2000 \times 200$. \emph{The loading factor was $1$}}\label{T:S1}
\begin{tabular}{|l|l|l|l|l|l|l|l|l|l|l|l|}
\hline
Model & Error  & r   & Rank PA & PA & Sign Flip & DPA & ACT & DDPA & $IC_1$ & $IC_2$ & $IC_3$ \\
\hline
PCA & Normal & 12   & 1 & 1  & 0 & 0.8 & 1 & 0.81 & 0 & 0 & 0\\
PCA & Normal  & 24  & 0.98  & 0.98  & 0 & 0 & 0.98 & 0 & 0 & 0 & 0\\
PCA & Normal &  48  & 0.11 & 0.12 & 0 & 0 & 0.1 & 0 & 0 & 0 & 0\\ 
\hline
FA & Normal & 12  & 1  & 1 & 0.01 & 0.8 & 1 & 0.8 & 0 & 0 & 0 \\
FA & Normal & 24 & 0.99 & 0.99 & 0 & 0 & 0.99 & 0 & 0 & 0 & 0\\
FA & Normal & 48  & 0.07 & 0.07 & 0 & 0 & 0.07 & 0 & 0 & 0 & 0\\
\hline 
PCA & t & 12   & 1 & 0.1  & 0 & 0  & 0 & 0 & 0 & 0 & 0\\
PCA & t  & 24  & 0.98  & 0.09  & 0 & 0 & 0 & 0 & 0 & 0 & 0\\
PCA & t &  48  & 0.08 & 0.18 & 0.06 & 0 & 0.12 & 0 & 0 & 0 & 0\\ 
\hline
FA & t & 12  & 1  & 0.15  & 0 & 0 & 0.12 & 0 & 0 & 0 & 0\\
FA & t & 24 & 0.99 & 0.03 & 0 & 0 & 0 & 0 & 0 & 0 & 0\\
FA & t & 48  & 0.1 & 0.22 & 0.03 & 0 & 0.15 & 0 & 0 & 0 & 0\\
\hline
\end{tabular}
\end{table}

\begin{table}
\centering
\caption{High Signal Mixture Scenario: The accuracy comparison between different methods for estimating the number of components in PCA and FA when the data matrix was of the size $2000 \times 200$. \emph{The loading factor was $20$}.}\label{T:S2}
\begin{tabular}{|l|l|l|l|l|l|l|l|l|l|l|l|}
\hline
Model & Error Dist.  & r  & Rank PA & PA & Sign Flip & DPA & ACT & DDPA &$IC_1$ & $IC_2$ & $IC_3$ \\
\hline
PCA & Normal & 12   & 1 & 1  & 0 & 0.77 & 1 & 0.75 & 1 & 1 & 1\\
PCA & Normal  & 24  & 0.99  & 0.99  & 0.02 & 0 & 0.99 & 0 & 0 & 0 & 0\\
PCA & Normal &  48  & 0.11 & 0.11 & 0 & 0 & 0.11 & 0 & 0 & 0 & 0\\ 
\hline
FA & Normal & 12  & 1  & 1 & 0 & 0.79 & 1 & 0.85 & 0.54 & 0.57 & 0.15\\
FA & Normal & 24 & 0.97 & 0.97 & 0.02 & 0 & 0.97 & 0 & 0 & 0 & 0\\
FA & Normal & 48  & 0.18 & 0.18 & 0 & 0 & 0.18 & 0 & 0 & 0 & 0\\
\hline 
PCA & t & 12   & 1 & 1  & 0.01 & 0  & 1 & 0 & 0.76 & 0.76 & 0.68\\
PCA & t  & 24  & 1  & 1  & 0 & 0 & 1 & 0 & 0 & 0 & 0\\
PCA & t &  48  & 0.16 & 0.16 & 0 & 0 & 0.16 & 0 & 0 & 0 & 0\\ 
\hline
FA & t & 12  & 1  & 1  & 0 & 0 & 1 & 0 & 0.02 & 0.05 & 0\\
FA & t & 24 & 0.99 & 0.99 & 0.01 & 0 & 0.99 & 0 & 0 & 0 & 0\\
FA & t & 48  & 0.2 & 0.2 & 0 & 0 & 0.2 & 0 & 0 & 0 & 0\\
\hline
\end{tabular}
\end{table}

\begin{table}
\centering
\caption{Low Signal Scenario: The accuracy comparison between different methods for estimating the number of components in PCA and FA when the data matrix was of the size $2000 \times 400$. \emph{The loading factor was $1$}}\label{T:S3}
\begin{tabular}{|l|l|l|l|l|l|l|l|l|l|l|l|}
\hline
Model & Error & r   & Rank PA & PA & Sign Flip & DPA & ACT & DDPA & $IC_1$ & $IC_2$ & $IC_3$  \\
\hline
PCA & Normal & 12   & 1 & 1  & 0 & 1 & 1 & 1 & 0 & 0 & 0\\
PCA & Normal  & 24  & 1  & 1  & 0 & 0.84 & 1 & 0.84 & 0 & 0 & 0\\
PCA & Normal &  48  & 0.78 & 0.82 & 0 & 0 & 0.97 & 0 & 0 & 0 & 0\\ 
\hline
FA & Normal & 12  & 1  & 1 & 0.02 & 1 & 1 & 1 & 0 & 0 & 0\\
FA & Normal & 24 & 1 & 1 & 0 & 0.81 & 1 & 0.83 & 0 & 0 & 0\\
FA & Normal & 48  & 0.73 & 0.76 & 0 & 0 & 0.96 & 0 & 0 & 0 & 0\\
\hline 
PCA & t & 12   & 1 & 0  & 0 & 0  & 0 & 0 & 0 & 0 & 0\\
PCA & t  & 24  & 1  & 0  & 0 & 0 & 0 & 0 & 0 & 0 & 0\\
PCA & t &  48  & 0.83 & 0 & 0 & 0 & 0 & 0 & 0 & 0 & 0\\ 
\hline
FA & t & 12  & 1  & 0  & 0 & 0 & 0 & 0 & 0 & 0 & 0\\
FA & t & 24 & 1 & 0 & 0 & 0 & 0 & 0 & 0 & 0 & 0\\
FA & t & 48  & 0.8 & 0 & 0 & 0 & 0 & 0 & 0 & 0 & 0\\
\hline
\end{tabular}
\end{table}

\begin{table}
\centering
\caption{High Signal Scenario: The accuracy comparison between different methods for estimating the number of components in PCA and FA when the data matrix was of the size $2000 \times 400$. \emph{The loading factor was $20$}.}\label{T:S4}
\begin{tabular}{|l|l|l|l|l|l|l|l|l|l|l|l|}
\hline
Model & Error & r  & Rank PA & PA & Sign Flip & DPA & ACT & DDPA & $IC_1$ & $IC_2$ & $IC_3$  \\
\hline
PCA & Normal & 12   & 1 & 1  & 0 & 0.99 & 1 & 1 & 1 & 1 & 1\\
PCA & Normal  & 24  & 1  & 1  & 0 & 0.78 & 1 & 0.75 & 0 & 0 & 0\\
PCA & Normal &  48  & 0.96 & 0.96 & 0 & 1 & 0.96 & 0 & 0 & 0 & 0\\ 
\hline
FA & Normal & 12  & 1  & 1 & 0 & 1 & 1 & 1 & 1 & 1 & 1 \\
FA & Normal & 24 & 1 & 1 & 0 & 0.85 & 1 & 0.8 & 0 & 0 & 0\\
FA & Normal & 48  & 0.96 & 0.96 & 0 & 0 & 0.99 & 0 & 0 & 0 & 0\\
\hline 
PCA & t & 12   & 1 & 1  & 0 & 0  & 1 & 0 & 0.62 & 0.64 & 0.37\\
PCA & t  & 24  & 1  & 1  & 0 & 0 & 1 & 0 & 0 & 0 & 0\\
PCA & t &  48  & 0.99 & 0.98 & 0 & 0 & 0.97 & 0 & 0 & 0 & 0\\ 
\hline
FA & t & 12  & 1  & 1  & 0 & 0 & 1 & 0 & 0.36 & 0.41 & 0.06\\
FA & t & 24 & 1 & 1 & 0 & 0 & 1 & 0 & 0 & 0 & 0\\
FA & t & 48  & 0.99 & 0.99 & 0.01 & 0 & 0.98 & 0 & 0 & 0 & 0\\
\hline
\end{tabular}
\end{table}

We also consider the performance of all the methods on the scenario where the loading factor is not the same for all the factors, i.e., a mixture scenario. In table \ref{T:S5}, for the mixture of low loadings we notice that PA's accuracy is zero for the number of factors varied over $\{12, 24\}$. Bai and Ng's $IC_1, IC_2, IC_3$ does better when the number of factors is $12$. We also see that DPA and DDPA do better in one scenario when the number of factor was $24$. However, it is easy to see that rank RPA is the most robust of all the methods. 
\begin{table}[h]
\centering
\caption{Low Loading Scenario: The accuracy comparison between different methods for estimating the number of components in PCA and FA when the data matrix was of the size $2000 \times 400$. \emph{The loading factor was a mixture of 1 and 6.}}\label{T:S5}
\begin{tabular}{|l|l|l|l|l|l|l|l|l|l|l|l|}
\hline
Model & Error & r & Rank PA & PA & Sign Flip & DPA & ACT & DDPA & $IC_1$ & $IC_2$ & $IC_3$  \\
\hline
PCA & Normal & 12   & 1 & 0  & 0 & 1 & 1 & 0.99 & 1 & 1 & 1\\
PCA & Normal  & 24  & 1  & 0  & 0 & 0.82 & 1 & 0.83 & 0 & 0 & 0\\
\hline
FA & Normal & 12  & 1  & 0 & 0 & 0.99 & 0.53 & 0.92 & 1 & 1 & 1 \\
FA & Normal & 24 & 0 & 0 & 0.01 & 0.86 & 0 & 0.52 & 0 & 0 & 0\\
\hline 
PCA & t & 12   & 1 & 0  & 0 & 0  & 0.01 & 0 & 0.61 & 0.65 & 0.45\\
PCA & t  & 24  & 1  & 0  & 0 & 0 & 0.02 & 0 & 0 & 0 & 0\\
\hline
FA & t & 12  & 0.92  & 0  & 0 & 0 & 0.52 & 0 & 0.51 & 0.54 & 0.31\\
FA & t & 24  & 0 & 0 & 0 & 0 & 0 & 0 & 0 & 0 & 0\\
\hline
\end{tabular}
\end{table}

\begin{table}
\centering
\caption{High Loading Scenario: The accuracy comparison between different methods for estimating the number of components in PCA and FA when the data matrix was of the size $2000 \times 400$. \emph{The loading factor was a mixture of 3 and 18.}}\label{T:S6}
\begin{tabular}{|l|l|l|l|l|l|l|l|l|l|l|l|}
\hline
Model & Error & r  & Rank PA & PA & Sign Flip & DPA & ACT & DDPA & $IC_1$ & $IC_2$ & $IC_3$  \\
\hline
PCA & Normal & 12   & 1 & 0  & 0 & 0.99 & 1 & 0.99 & 1 & 1 & 1\\
PCA & Normal  & 24  & 1  & 0  & 0.01 & 0.79 & 1 & 0.73 & 0 & 0 & 0\\
\hline
FA & Normal & 12  & 0.99  & 0 & 0 & 0.99 & 0.49 & 0.91 & 1 & 1 & 1 \\
FA & Normal & 24 & 0 & 0 & 0.02 & 0.79 & 0 & 0.7 & 0 & 0 & 0\\
\hline 
PCA & t & 12   & 1 & 0  & 0 & 0  & 0.99 & 0 & 0.56 & 0.65 & 0.33\\
PCA & t  & 24  & 1  & 0  & 0 & 0 & 0.98 & 0 & 0 & 0 & 0\\
\hline
FA & t & 12  & 0.98  & 0  & 0 & 0 & 0.37 & 0 & 0.43 & 0.48 & 0.15\\
FA & t & 24  & 0 & 0 & 0.01 & 0 & 0 & 0.02 & 0 & 0 & 0\\
\hline
\end{tabular}
\end{table}

\section{Real Data Analysis}\label{s.data}
For the real data analysis, we considered two popular datasets in the literature from psychology and economics for checking the real-world validity of our method. The real data analysis is as follows.
\subsection{Psychology Data}
The psychology data concerns the mental ability scores of the seventh and eighth grade students on the nine test questions originally collected by Holzinger and Swineford (1939) \cite{holzinger1939}. The nine test questions are based on visual perception, cubes, lozenges, paragraph comprehension, sentence completion, word meaning, speedy addition, speedy counting of dots, and speedy discrimination between straight and curved capitals and is widely used in the literature see Joreskog (1969) \cite{joreskog1969}.

\begin{table}[H]
\center
\caption{Estimated Number of Factors. For PA and Rank PA, we report the median number of factors that was estimated over $200$ MCMC simulations. }\label{T:FedMD}
\begin{tabular}{|l|l|l|l|l|l|l|l|}
\hline
Data & Rank PA & PA & DPA & ACT  & $IC_1$ & $IC_2$ & $IC_3$   \\
\hline
Holzinger and Swinefold & 3 & 3 & 3 & 3 & 3 & 3 & 3 \\
\hline
\end{tabular}
\end{table}

\subsection{Fred-MD Data}
The federal reserve bank of St-louis posts the monthly macro-economic data capturing the multiple facets of the US economy
 see McCracken and Ng (2015) \cite{fredmd2015}. The economic variables are typically transformed into stationary data by considering the differences, or by taking the differences of the logarithms, or by using the rate of change etc. The multiple variable that it tracks range into hundreds and require a factor analysis study for understanding the key macro-economic factors affecting the US economy. We considered the following data from the 1st of January 1959 to 1st August 2025 obtained from the federal reserve of St-louis website. For the analysis, we considered the full data consisting of $116$ columns over the data ranging from 1st January 1959 to 1st Dec 2006 (before the financial recession of 2007-08), the data after 1st January 2007 (during and after the financial recession data). On each of these versions, we ran the competing method along with ours to estimate the number of factors.
Table \ref{T:FedMD} lays out the result for the estimated number of the factors for the three periods. McCracken and Ng (2015) \cite{fredmd2015} find the estimated number of factors as eight. However, Bai and Ng (2002) \cite{bai2002determining} tend to be sensitive to the hyperparameter corresponding to the maximum number of factors. McCracken and Ng (2015) \cite{fredmd2015} found the first four factors to be stable (across databases) while factors from five to eight relative less stable (across databases). Apart from PA and Rank PA none of the other methods are close. However, Rank PA seems to be closer to the macro-economic view as it estimates $7$ number of factors for both the full data and post 2007 while it estimated a total of $6$ number of factors for before 2007 period. On the other hand, PA estimated $6$ as the number of factors for the during and after the financial recession. Unfortunately, other methods such as Bai's information criterion and DPA tend to be sensitive to the maximum number of factors and here with the maximum number of factors set as $20$ and $50$, they are overestimating the number of factors. 
Fan's ACT method is estimating the number of factors as $5$ for before 2007 series and after 2007 series unable to measure any change because of the 2007 financial crisis.
\begin{table}[H]
\center
\caption{Estimated Number of Factors. For PA and Rank PA, we report the median number of factors that was estimated over $200$ MCMC simulations. The maximum number of factors for Bai and Ng (2020) was $20$ while for DPA was $50$.}\label{T:FedMD}
\begin{tabular}{|l|l|l|l|l|l|l|l|}
\hline
Fed-Md Data & Rank PA & PA & DPA & ACT  & $IC_1$ & $IC_2$ & $IC_3$   \\
\hline
Before 2007 & 6 & 5  & 50 & 5  & 20  & 20 & 20  \\
\hline
Onwards 2007 & 7 & 6 & 50 & 5  & 20  & 20 & 20 \\
\hline
Full Data & 7 & 7  & 50  & 5 & 20  & 20  & 20 \\
\hline
\end{tabular}
\end{table}

\section{Discussion}\label{s.discussion}
We proposed a rank-based nonparametric version of parallel analysis called Rank PA. Interestingly, we discover that Rank PA is less sensitive to the distributional assumptions or the signal strength. Moreover, our method does not require any fourth moment condition for estimating the number of factors or number of components massively boosting the applicability of our approach to heavy-tailed distributions. Finally, we show that Rank PA can estimate the number of factors or components up to $O(n^{1/3 - \tau})$ under relaxed conditions compared to the other competing approach. The strength of Rank PA comes from the observation that the latent dimension of a data matrix is a subtle property of the order or the arrangement of its column entries. This arrangement is missed by all the methods that overtly rely on the likelihoods. A limitation of our approach is that it is tailor-made for continuous error distributions. A future direction for research would be to extend the above approach for discrete error distributions. The github repository for our work can be found at the url \url{https://github.com/chetkar/rankPA}.


\bibliography{reference}

@article{fan2019factor,
  title={Estimating Number of Factors by Adjusted Eigenvalues Thresholding},
  author={Fan, Jianqing and Guo, Jianhua and Zheng, Shurong},
  journal={Journal of the American Statistical Association},
  year={2020}
}

@article{guttman1954factor,
  title={Some necessary conditions for common-factor analysis},
  author={Guttman, Louis},
  journal={Psychometrika},
  volume={19},
  number={2},
  pages={149-161},
  year={1954},
}

@article{onatski2010factor,
  title={Determining the number of factors from empirical distribution of eigenvalues},
  author={Onatski, Alexei},
  journal={The Review of Economics and Statistics},
  volume={92},
  number={4},
  pages={1004-1016},
  year={2010},
}

@article{dobriban2018dpa,
  title={Deterministic parallel analysis: an improved method for selecting factors and principal components},
  author={Dobriban, Edgar and Owen, Art B},
  journal={Journal of the Royal Statistical Society. Series B (Methodological)},
  volume={81},
  number={1},
  pages={163-183},
  year={2019},
}

@article{dobriban2017permutation,
  title={Permutation methods for factor analysis and PCA},
  author={Dobriban, Edgar},
  journal={Annals of Statistics},
  volume={48},
  number={5},
  year={2020}
}

@article{buja1992,
  title={Remarks on parallel analysis},
  author={Buja, Andreas and Eyuboglu, Nermin},
  journal={Multivariate behavioral research},
  volume={27},
  number={4},
  pages={509-540},
  year={1992},
 }

@article{horn1965,
  title={A rationale and test for the number of factors in factor analysis},
  author={Horn, John L},
  journal={Psychometrika},
  volume={30},
  number={2},
  pages={179-185},
  year={1965},
 }

@article{bai2002determining,
  title={Determining the number of factors in approximate factor models},
  author={Bai, Jushan and Ng, Serena},
  journal={Econometrica},
  volume={70},
  number={1},
  pages={191--221},
  year={2002},
  publisher={Wiley Online Library}
}

@article{johnstone2001distribution,
  title={On the distribution of the largest eigenvalue in principal components analysis},
  author={Johnstone, Iain M},
  journal={Annals of Statistics},
    volume={29},
  number={2},
  pages={295--327},
  year={2001},
  publisher={JSTOR}
}

@article{baik2005phase,
  title={Phase transition of the largest eigenvalue for nonnull complex sample covariance matrices},
  author={Baik, Jinho and Arous, G{\'e}rard Ben and P{\'e}ch{\'e}, Sandrine},
  journal={Annals of Probability},
  volume = {33},
  number = {5},
  pages={1643--1697},
  year={2005},
  publisher={JSTOR}
}

@article{candes2011robust,
  title={Robust principal component analysis?},
  author={Cand{\`e}s, Emmanuel J and Li, Xiaodong and Ma, Yi and Wright, John},
  journal={Journal of the ACM (JACM)},
  volume={58},
  number={3},
  pages={11},
  year={2011},
  publisher={ACM}
}

@book{anderson1958introduction,
	Author = {Anderson, Theodore W},
	Publisher = {Wiley New York},
	Title = {An Introduction to Multivariate Statistical Analysis},
	Year = {2003}}

@article{decelle2011,
author={Decelle, Aurelien and Krzakala, Florent and Moore, Cristopher and Zdeboro\'{v}a, Lenka},
title={Asymptotic Analysis of the Stochastic Block Model for Modular Networks and its logarithmic applications},
journal={Physics Review E},
volume={84},
pages={066106},
year={2011}
}

@article{dobriban2017,
author={Dobriban, Edgar},
title={Permutation Methods for Factor Analysis and PCA},
journal={Preprint, https://arxiv.org/abs/1710.00479, to appear in the Annals of Statistics},
year={2017}
}

@article{cattell1966,
author={Cattell, Raymond B.},
title={The Scree Test for the Number of Factors},
journal={Multivariate Behavioral Research},
volume={1},
number={2},
year={1966}
}

@article{kaiser1960,
author={Kaiser, H. F.},
title={The application of electronic computers to factor analysis},
journal={Educational and Physiological Measurement},
volume={20},
year={1960},
pages={141--151}
}

@article{jolliffe2016,
author={Jolliffe, Ian T. and Cadima, Jorge},
title={Principal Component Analysis : A review and recent developments},
journal={Philosophical Transactions of the Royal Society A},
volume={374},
number={2065},
year={2016}
}

@article{ke2020ci,
author={Ke, Zheng Tracy and Ma, Yucong and Lin, Xihong},
title={Estimation of the number of spiked covariance matrix by bulk eigenvalue matching analysis},
journal={preprint, https://arxiv.org/pdf/2006.00436.pdf},
year={2020}
}

@article{bai2002factor,
author={Bai, J. and Ng, S.},
title={Determining the number of factors in approximate factor models},
journal={Econometrica},
volume={70},
number={1},
pages={191--221},
year={2002}
}

@article{fan2008factor,
author={Fan, Jianqing and Fan, Yingying and Lv, Jinchi},
title= {High dimensional covariance matrix estimation using a factor model},
journal={Journal of Econometrics},
volume={147},
number={1},
pages={186-197},
year={2008}
}

@article{fan2020factor,
author={Fan, J. and Guo, J. and Zheng, S.},
title={Estimating number of factors by adjusted eigenvalues thresholding},
journal={Journal of the American Statistical Association (to appear)},
year={2020}
}

@article{bai2008factor_review,
author={Bai, J. and Ng, S. },
title={Large Dimensional Factor Analysis},
journal={Foundations and Trends in Econometrics},
volume={3},
number={2},
year={2008},
pages={89--163}
}

@article{costello2005,
author={Costello, Anna B. and Osborne, Jason},
title={Best practices in exploratory factor analysis : four recommendations for getting the most from your analysis},
journal={Practical Assessment, Research, and Evaluation},
volume={10},
year={2005}
}

@article{onatski2009,
author={Onatski, Alexei},
title={Testing Hypotheses about the number of Factor in Large Factor Models},
journal={Econometrica},
volume={77},
number={5},
pages={1447--1479},
year={2009}
}

@article{jha2023,
author={Jha, Chetkar and Barnett, Ian},
title={Confidence Interval of the Number of Components in Factor Analysis and PCA via Subampling},
journal={https://arxiv.org/pdf/2205.04945.pdf},
year={2022}
}

@book{brown2014,
author={Brown, T. A.},
title={Confirmatory Factor Analysis for Applied Research},
publisher={Guilford, New York},
year={2014}
}

@article{spearman1904,
author={Spearman, C.},
title={"General intelligence", objectivity determined and measured},
journal={American Journal of Psychology},
volume={15},
pages={201--292},
year={1904}
}

@article{randomsignflips,
author={Hong, David and Sheng, Yue and Dobriban, Edgar},
title={Selecting the number of components via random signflips},
journal={https://arxiv.org/abs/2012.02985},
year={2020}
}

@article{pearson1901,
author={Pearson, Karl},
title={On Lines and Planes of Closest Fit to Systems of Points in Space},
journal={The London, Edinburgh, and Dublin Philosophical Magazine and Journal of Science},
volume={2},
number={11},
pages={559--572},
year={1901}
}

@article{hotelling1933,
author={Hotelling, H},
title={Analysis of a complex of statistical variables into principal components},
journal={Journal of Educational Psychology},
volume={24},
number={6},
pages={417--441},
year={1933}
}

@book{vom,
author ={Thurstone, L},
title={The Vector of Mind. Multiple-factor analysis for the isolation of primary traits},
publisher={University of Chicago},
year={1935}
}

@article{siluta2023,
    author = {Szafraniec-Siluta, Ewa and Ardan, Roman and Strzelecka, Agnieszka and Zawadzka, Danuta},
    title = {Application of factor analysis to reduce the dimensionality of the determinants of equity capital return on European Union farms},
    journal ={Procedia Computer Science} ,
    volume={225},
    year = {2023},
    pages={4433--4442}
}

@article{cekic2022,
author={Skaka-Čekić, F. and Baraković Husić and J., Odžak, A. et al.},
title={Dimensionality reduction of independent influence factors in the objective evaluation of quality of experience}, 
journal={Scientific Reports},
volume={12}, 
year={2022}
}

@book{jolliffe2002,
author={Jolliffe, Ian},
title={Principal Component Analysis},
publisher={Springer},
year={2002}
}

@article{huang2008,
author={Shen, Haipen and Huang, Jianhua Z.},
title={Sparse Principal Components via Regularized Low Rank Approximation},
journal={Journal of Multivariate Analysis},
volume={99},
number={6},
pages={1015--1034},
year={2008}
}

@article{witten2009,
author={Witten, Daniela M. and Tibshirani, Robert and Hastie, Trevor},
title={A penalized decomposition with applications to sparse principal components and canonical correlation analysis},
journal={Biostatistics},
volume={10},
number={3},
pages={515--534},
year={2009}
}

@article{journee2010,
author={Journee, Michel and Nesterov, Yurii and Richtarik, Peter and Sepulchre, Rodolphe},
title={Generalized power method for sparse principal component analysis},
journal={Journal of Machine Learning Research},
volume={11},
pages={517--553},
number={2},
year={2010}
}

@article{huang2019,
author={Huang, Wen and Wei, Ke},
title={Extending fista to riemannian optimization for sparse pca},
journal={Technical Reports},
year={2019}
}

@article{bertsimas2022,
author={Bertsimas, Dimitris and Kitane, Driss Lahlou},
title={Sparse PCA: A Geometric Approach},
journal={Journal of Machine Learning Research},
pages={1--33},
year={2022}
}

@book{cattellfactor1988,
author={Nesselroade, John R. and Cattell, R.B.},
title={Handbook of Multivariate Experimental Psychology. Perspectives on Individual Differences.},
publisher={Springer, Boston, MA},
year={1988}
}

@article{elhaik2022,
author={Elhaik, Eran},
title={Principal Component Analyses (PCA)-based findings in population genetic studies are highly biased and must be reevaluated},
journal={Nature},
year={2022}
}

@article{edwards1963,
author={Edwards, A. and Cavalli-Sfoza, L. L.},
title={Analysis of human evolution},
journal={Genetics Today, Proceedings of 11th International Congress of Genetics},
volume={3},
pages={923--933},
year={1963}
}

@article{wirth2016,
author={Wirth, R. J. and Edwards, Michael C.},
title={Item Factor Analysis: Current Approaches and Future Directions},
journal={Psychology Methods},
pages={58--79},
year={2011}
}

@article{covidfactor2025,
author={Dalawi, I and Isa, M. R. and Aimran, N.},
title={Exploratory factor analysis on the development and validation of the understanding, attitude, practice and health literacy questionnaire on COVID-19 in Malay language},
journal={Sci Rep 15},
year={2025}
}

@article{schnatter2025,
author={Fr\"{u}hwirth-Schnatter, Sylvia and Hosszejni, Darjus and Lopes, Hedibert Freitas},
title={Sparse Bayesian Factor Analysis When the Number of Factors is Unknown},
journal={Bayesian Analysis},
volume={20},
number={1},
pages={213--344},
year={2025}
}

@article{owen16,
author={Owen, Art B.and Wang, Jingshu},
title={Bi-Cross-Validation for Factor Analysis},
journal={Statistical Science},
volume={31},
number={1},
pages={119--139},
year={2016}
}

@article{zhou2023,
author={Zhou, Y. and Luo, K. and Liang, L. et al.},
title={A new Bayesian factor analysis method improves detection of genes and biological processes affected by perturbations in single-cell CRISPR screening},
journal={Nat Methods},
volume={20},
pages={1693--1703},
year={2023}
}

@article{bfactor2014,
author={Conti, Gabriella and Fr\"{u}hwirth-Schnatter, Sylvia and Heckman, James J. and Piatek, R\'{e}mi},
title={Bayesian exploratory factor analysis},
journal={Journal of Econometrics},
volume={183},
number={1},
year={2014}
}

@book{lopes2014,
author={Lopes, Hedibert Freitas},
title={Bayesian Inference in the Social Sciences},
publisher={Wiley},
year={2014}
}

@article{west2004,
author={Lopes, Hedibert Freitas and West, Mike},
title={Bayesian Model Assessment In Factor Analysis},
journal={Statistical Sinica},
volume={14},
pages={41--67},
year={2004}
}

@article{carvalho2008,
author={Carvalho, Carlos M. and Chang, Jeffrey and Lucas, Joseph E. and Nevins, Jospeh R. and Wang, Quanli and West, Mike},
title={High-Dimensional Sparse Factor Modeling: Application in Gene Expression Genomics},
journal={Journal of the American Statistical Association},
volume={103},
number={484},
pages={1438--1456},
year={2008}
}

@article{knowles2011,
author={Knowles, David and Ghahramani, Zoubin},
title={Nonparametric Bayesian Sparse Factor Models With Application To Gene Expression Modeling},
journal={Annals of Applied Statistics},
volume={5},
number={2B},
pages={1534--1552},
year={2011}
}

@article{watson2002,
    author={Stock, J. H. and  Watson, M. W.},
    title={Forecasting Using Principal Components from a Large Number of Predictors},
    journal={Journal of the American Statistical Association}, volume={97},
   number={460}, 
   pages={1167--1179},
    year={2002}
}

@article{LI201776,
author={Hongjun, Li and Qi, Li and Yutang, Shi},
title = {Determining the number of factors when the number of factors can increase with sample size},
journal = {Journal of Econometrics},
volume = {197},
number = {1},
pages = {76-86},
year = {2017}
}

@article{caro2025,
author={Caro, Angela and P\~{e}na, Daniel},
title={Selecting the Number of Factors in Multivariate Time Series},
journal={Journal of Time Series},
volume={46},
number={1},
year={2025}
}

@article{haslbeck2024,
author= {Haslbeck, J. M. B. and van Bork, R.},
title={Estimating the number of factors in exploratory factor analysis via out-of-sample prediction errors},
journal={Psychological Methods},
volume={29},
number={1}, 
pages={48--64},
year={2024}
}

@article{schwartz1978,
author={Schwartz, Gideon},
title ={Estimating the Dimension of a Model},
journal ={Annals of Statistics},
volume={6},
number={2},
year={1978}
}

@article{joreskog1969,
author={J\"{o}reskog, K. G.},
title={A General Approach to Confirmatory Maximum Likelihood Factor Analysis},
journal={Psychometrika},
pages={183--202},
year={1969}
}

@article{holzinger1939,
author={Holzinger, K. and Swineford, F.},
title={A study in factor analysis: The stability of a bifactor solution. Supplementary Educational Monograph},
volume={48},
journal={University of Chicago Press},
year={1939}
}

@article{fredmd2015,
author= {McCracken, M. W. and Ng, S.}, 
title={FRED-QD: A Quarterly Database for Macroeconomic Research}, 
journal={working papers 2015-012B, Federal Reserve Bank of St. Louis},
year={2015}
}

\end{document}